\documentclass{iopart}
\usepackage{graphicx}

\begin{document}

\title{Anomalous Longitudinal Magnetic Field near the Surface of Copper Conductors}
\author{S.~Kraft, A.~G\"{u}nther, H.~Ott, D.~Wharam*,
C.~Zimmermann, and J.~Fort\'{a}gh}

\address{Physikalisches Institut der Universit\"{a}t
T\"{u}bingen, Auf der Morgenstelle 14, 72076 T\"{u}bingen,
Germany}

\address{* Institut f\"{u}r Angewandte Physik der
Universit\"{a}t T\"{u}bingen, Auf der Morgenstelle 10, 72076
T\"{u}bingen, Germany}

\begin{abstract}
We have used ultracold atoms to characterize the magnetic field
near the surface of copper conductors at room temperature carrying
currents between 0.045~A and 2~A. In addition to the usual
circular field we find an additional, $1000- 10000$ times smaller
longitudinal field. The field changes its strength periodically
with a period of $200-300\,\mu$m.
\end{abstract}

Recently, several experiments have been successful in bringing
clouds of ultracold alkali atoms close to the surface of copper
current conductors \cite{surface experiments}. The clouds have
been prepared below and above the critical temperature for
Bose-Einstein condensation and in all cases a fragmentation of the
atomic distribution has been observed which suggests the presence
of an as yet unexplained potential caused by the conductors. In
this letter we show that this potential is of magnetic origin and
is due to a longitudinal field component which is three to four
orders of magnitude smaller than the usual circular field of the
conductor. This result demonstrates that ultracold atoms, as
provided by a Bose-Einstein condensate can be used as a sensitive
probe for magnetic fields. The probing principle is based on the
force that is acting on the atoms in a static magnetic field. It
is proportional to the gradient of the magnetic field modulus
\cite{traps} and results in a change of the atomic distribution
which can be imaged by standard techniques. The spatial dependence
of the magnetic field modulus can be probed by moving the atomic
cloud within the magnetic field. This is possible either by using
optical forces \cite{tweezers} or by means of magnetic potentials.
In the latter case, the sample field, which is to be measured, and
the trapping field may superimpose. This complicates the analysis,
however, it can also be used to separate different components of
the sample field.

\begin{figure} \centering
\includegraphics[height=4cm]{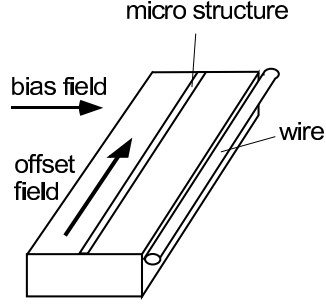}
\caption{\label{fig1} Trap setup. Two conductors are used: a
microfabricated copper conductor path with a width of 30~$\mu$m
and an ordinary copper wire with a diameter of 90~$\mu$m. Together
with a perpendicular bias field a waveguide is formed. The offset
field generates a non vanishing homogeneous field component in the
direction of the conductors.}
\end{figure}
Fig.~\ref{fig1} shows the setup of our experiment. The field
configuration is suitable for selectively observing a longitudinal
magnetic field component. In addition to the field generated by
the conductor, two homogenous fields are present, one oriented
perpendicular to the wire (bias field) and the other oriented
parallel to the wire (offset field). The bias field
$B_\mathrm{bias}$ together with the usual circular field of the
conductor form a linear magnetic quadrupole field with a vanishing
magnetic field along a line parallel to the wire and separated
from its centre by
$d=\frac{\mu_0}{2\pi}\frac{I}{B_\mathrm{bias}}$. Here, $I$ is the
current in the conductor. Perpendicular to the line of vanishing
magnetic field its modulus increases linearly with the distance
and forms a waveguide like trapping potential for ultracold atoms
\cite{traps}. While the transverse motion of the atoms is confined
by the potential, the atoms can move freely along the longitudinal
direction. A hypothetical longitudinal field component would
generate an additional longitudinal potential that can be detected
by observing the atomic distribution along the wire. In fact, such
an unexpected field component can be observed in our experiments.
It is 3 to 4 orders of magnitude smaller than the usual circular
field but still strong enough to  trap the atoms in the
longitudinal direction. The atomic distribution shows a pronounced
periodicity of $200-300\,\mu$m which corresponds to a similar
modulation of the anomalous field component at the location of the
atoms.

We have used rubidium atoms to study the magnetic field of two
different types of copper conductor. The first conductor has a
rectangular cross section with a height of $2\,\mu$m and a width
of $30\,\mu$m. It has been electroplated onto an aluminium oxide
substrate as described in detail elsewhere
\cite{MakingMicrostructure}. It is operated with a current between
45~mA and 0.5~A corresponding to a current density between
$7.5\cdot 10^{4} \mathrm{\frac{A}{cm^{2}}}$ and $8.3\cdot
10^{5}\mathrm{\frac{A}{cm^{2}}}$, respectively. The bias field
amounts between 2~G and 22~G. The second conductor is an ordinary
copper wire as used for electronic circuits with a diameter of
$90\,\mu$m. It carries 0.3~A to 2~A and is operated with the same
bias field. To initially trap and cool the atomic cloud we apply
an offset field along the longitudinal direction that is slightly
inhomogeneous and forms a parabolic trapping potential in which
the atoms oscillate with a frequency of 14~Hz \cite{BECdetails}.
Here, a cloud of $10^{5}$ atoms is prepared at a temperature of
about $1\,\mu$K. The trapping and cooling procedure is described
elsewhere \cite{BECTuebingenPRL}. By turning off the longitudinal
trapping potential within 400~ms and keeping only a homogenous
offset field of 1.3~G, the atoms are released into the wave guide
for 100~ms where they are exposed to the anomalous longitudinal
field generated by the wire or by the microstructure,
\begin{figure} \centering
\includegraphics[width=7.5cm]{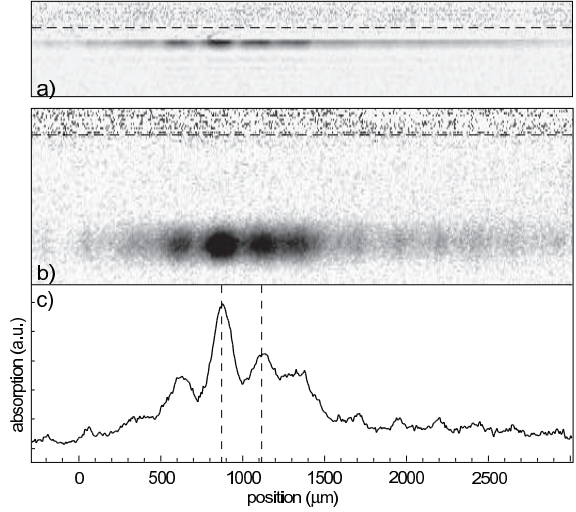}
\caption{\label{fig2} In a waveguide near the copper conductor the
density distribution of the atom cloud shows a periodic
fragmentation. a) shows the cloud in the initial waveguide. The
dashed line indicates the surface of the microstructure. To
prepare the atoms in this trap the axial confinement was ramped
down in 400 ms. b) shows the atoms after 10~ms time of flight. The
periodic structure appeares more clearly. In c) a integrated scan
of b) is plotted.}
\end{figure}
respectively. Fig.~\ref{fig2}a shows the density distribution of
the atoms near the surface of the microstructure at a current of
0.045~A and a bias field of 2~G. Better imaging is possible 10~ms
after all magnetic fields have been turned off (Fig.~\ref{fig2}b).
In this time the atoms have been separated from the surface while
falling under gravity. We observe a spatial modulation of the
density distribution with a periodicity of $260\pm 15\,\mu$m.
Similar properties are found for the wire (Fig.~\ref{fig3}.),
\begin{figure} \centering
\includegraphics[width=7.5cm]{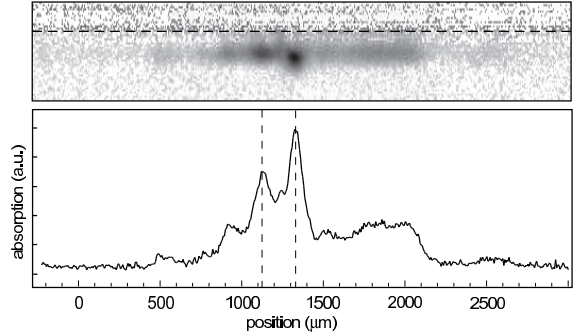}
\caption{\label{fig3} The periodic structure at the 90 $\mu$m
copper wire after 5~ms time of flight. The current in the wire,
the bias field, and the offset field are 0.2~A, 6~G and 2.6~G
respectively.
}
\end{figure}
however with less pronounced periodicity. Nevertheless, the
fourier spectrum of the density distribution shows a clear peak at
$220\,\mu$m.

The magnetic origin of the anomalous surface potential can be
shown by changing the orientation of the offset field
$B_\mathrm{off}$. The atomic density distribution is now inverted
with the maxima transformed into minima and vice versa
\begin{figure} \centering
\includegraphics[width=7.5cm]{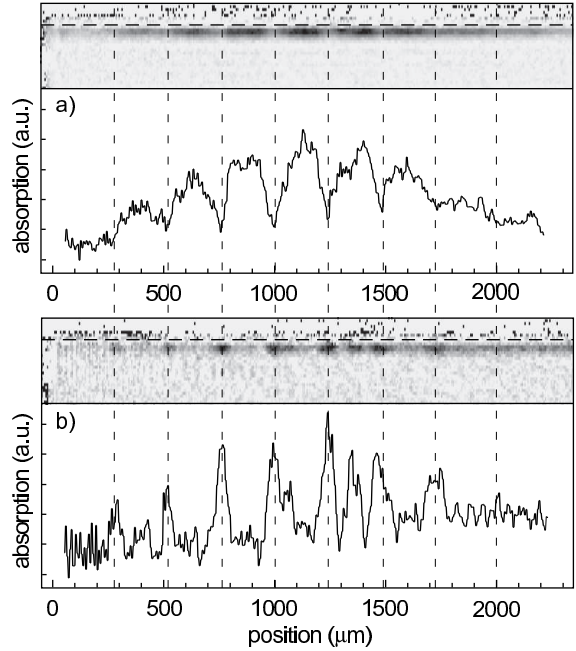}
\caption{\label{fig4} The position of the potential minima change
when the direction of the offset field is changed. In a) the atoms
are prepared in a waveguide by ramping down the axial confinement
in 400 ms. The current in the microfabricated conductor is
0.045~A. Together with a bias field of 2~G a waveguide is formed
at a distance of 45~$\mu$m from the surface. The offset field
parallel to the conductor is 1.3~G. The vertical dashed lines mark
the position of the potential maxima. In b) the atoms are
initially prepared in the same trap as in a). Next the orientation
of the offset field is flipped within 1~ms. After further 50~ms
the atoms are located in the position of the former potential
maxima.}
\end{figure}
(Fig.~\ref{fig4}). To explain this effect, we assume an anomalous
magnetic field $B_\mathrm{z}(z)$ and add it to $B_\mathrm{off}$.
This leads to a total field given by $\pm
B_\mathrm{off}+B_\mathrm{z}(z)$ with the sign of $B_\mathrm{off}$
depending on the orientation of the offset field. The resulting
potential is proportional to the modulus of the total field
$\left| \pm B_\mathrm{off}+B_{z}(z)\right|$ which is equivalent to
$\left| B_\mathrm{off}\pm B_{z}(z)\right| $. Therefore, flipping
the offset field has the same effect as inverting the surface
potential. In the case of a non magnetic potential, the potential
would simply add to the magnetic potential that is generated by
the offset field and inverting its orientation would have no
effect. The same applies for any magnetic field component which
does not have the same direction as the offset field. If
$B_\mathrm{z}(z)$ consists of a constant field $B_\mathrm{0}$ and
a modulated field $B_\mathrm{mod}(z)$, the constant part adds to
$B_\mathrm{off}$ with opposite sign in the two cases. However, the
change in sign of $B_\mathrm{mod}(z)$ still transforms minima into
maxima and vice versa. To investigate the origin of the modulated
field we have changed the orientation of the current in the
conductor (and the bias field) while holding the offset field
constant. Again the minima where transformed into maxima and vice
versa. This shows, that the anomalous magnetic field is caused by
the current in the wire rather than by permanent magnetic
inhomogeneities of the conductor. To complete this type of
experiment we flipped the orientation of all elements involved:
the current orientation in the conductor, the bias field and the
offset field. Then the original distribution was restored. The
observed dependence on the current provides a simple method to
eliminate the structuring effect of the surface potential on the
cloud. By periodically inverting the current and the bias field on
a fast time scale the atoms experience only a time avaraged
potential with a strongly reduced spatial dependence.

In order to estimate the strength of the anomalous magnetic field
component we further cool the atomic cloud below the critical
temperature where it undergoes transition into a Bose-Einstein
condensate \cite{BECsufacePRA}. In contrast to the experiments
described above we now keep the atoms trapped longitudinally. By
changing the bias field the condensate can be shifted relative to
the surface of the conductor. At large distances from the surface
the anomalous surface potential is small and does not affect the
condensate. However, at a critical distance the condensate splits
into two components. This occurs if the centre of the external
harmonic trap is placed at the position of a maximum of the
surface potential. Then the total potential shows a double well
structure with a barrier height that is comparable with the depth
of the surface potential. In fact, for a sinosoidal surface field
with a period of $220\,\mu$m and an amplitude of
$B_\mathrm{mod}(z)$, superimposed with the field of the harmonic
trap of $B_\mathrm{trap}(z)=84\,\mathrm{G/cm^2}\cdot z^{2}$ one
finds a ratio between the amplitude and the barrier hight of 2.
The barrier can split the condensate if its height exceeds the
chemical potential of the condensate. Since the chemical potential
can easily be measured by standard time of flight methods
\cite{chmPotMess} it provides a good estimate for the depth of the
combined potential. This value represents a lower bound for the
depth of the surface potential. A deviation of a pure sinosoidal
shape can result in a change of the potential depth in one order
of magnitude. For the wire with a current of 0.9~A, a bias field
of 16.5~G and an offset field of 2.6~G we observe the splitting of
the condensate at a distance of $109\,\mu$m from the surface. The
chemical potential is determined to be 250~nK resulting in an
estimation for the minimum depth of the surface potential of
$3.4\cdot 10^{-30}$~J. This corresponds to a magnetic field of
3.7~mG which is more than two orders of magnitude smaller than the
field strength of the circular field component of 16~G at that
distance. Qualitatively, the potential depth increases
monotonically  with the current, and decreases with the distance
to the surface.

We can further extend the analysis if we assume the Ansatz
$U=\tilde{C}\cdot I\cdot d^{-q}$ for the strength of surface
potential $U$ at a distance $d$ form the surface and a current $I$
in the conductor. Here, $\tilde{C}$ is a constant only depending
on $z$ and $q$ is a real number. To find the value of $q$ we have
determined the distance at which the two condensate parts are
separated by $80\,\mu$m. For five currents in the wire ranging
from 0.3~A to 1.2~A this distance increases from $80\,\mu$m to
$109\,\mu$m. For each data pair we have also determined the
chemical potential of the condensate and used it as a measure for
the height of the potential barrier. Since, to a good
approximation, the potential barrier is proportional to the depth
of the surface potential we can use the above Ansatz to describe
the data. The $\chi ^{2}$-parameter of a least square fit for
$q=1$, $q=2$, and $q=3$, with varying $\tilde{C}$ amounts to
$140$, $5.5$, and $11$, respectively. Best agreement is found for
$q=2.2$  with a $\chi^2$-parameter of 5.49, giving a preference to
a value around 2.

Repeating the experiment with the microstructure leads to a
different constant $\tilde{C}$. This difference can mostly be
eliminated by replacing the current $I$ with the product of the
current density $j$ and the width $w$ of the conductor. In the
case of the wire the radius is taken for the width. Then one
arrives at a heuristic expression for $U=C\cdot j\cdot w\cdot
d^{-2} $, with $C\simeq 5\cdot 10^{-42}\mathrm{m^{3}J/A}$. This
suggests that the field is generated at the surface rather than
inside the conductor since the product $j\cdot w$ may be
interpreted as being proportional to the amount of surface current
as ``seen'' by the atoms. Furthermore, the data recently reported
from other experimental research groups seem to fit with the
expression \cite{otherexperiments}. It is not possible to find a
similar expression with $d$ interpreted as the distance to the
centre of the conductor. Such an approach does not allow a
consistent description of all known observations.

For the development of a model that describes the physical origin
of the anomalous field it will be important to explain the
observed periodicity of several $100\,\mu$m of the anomalous field
component. It is most clearly observed in the microfabricated wire
and seems to be almost independent on the geometry of the
conductor. One may speculate that the fabrication process plays a
role. Imperfections in the conductor may inhibit or destroy the
periodicity. Candidates for a possible explanation are kink
instabilities of the current as known from plasma physics. Even
more speculative are current induced spin orientations at the
surface. Important experimental data are to be expected soon from
experiments carried out with gold conductors \cite{gold}.

\ack We thank E. Hinds, W. Ketterle, D. Kielpinski, J. Reichel, L.
Feenstra, G. Mih\'{a}ly and R. P. H\"ubener for valuable
discussions. This work was supported in part by the Deutsche
Forschungsgemeinschaft under Grant No.~Zi 419/3.

\end{document}